# Funding acknowledgment analysis: Queries and Caveats

Li Tang[1], Guangyuan Hu[2], Weishu Liu[3,*]

1. School of International Relations and Public Affairs, Fudan University, Shanghai 200433, China, Email: tang006@gmail.com

2. School of Public Economics and Administration, Shanghai University of Finance and Economics, Shanghai 200433, China, Email: hu.guangyuan@shufe.edu.cn

3. Corresponding author, Antai College of Economics and Management, Shanghai Jiao Tong University, Shanghai 200052, China, Email: wsliu08@163.com

Thomson Reuters' Web of Science (WoS) began systematically collecting acknowledgment information in August 2008. Since then, bibliometric analysis of funding acknowledgment (FA) has been growing and has aroused intense interest and attention from both academia and policy makers. Examining the distribution of FA by citation index database, by language, and by acknowledgment type, we noted coverage limitations and potential biases in each analysis. We argue that in spite of its great value, bibliometric analysis of FA should be used with caution.

## Introduction

Acknowledgments in scientific publications express authors' gratitude to diverse entities who funded, inspired, or contributed to their research (Cronin et al., 1993; Costas & Leeuwen, 2012; Salager-Meyer et al., 2011; Tiew & Sen, 2002). Though they were once called "long neglected textual artefacts" (Cronin et al., 1992), acknowledging support is becoming standard practice in scientific communications (Cronin, 2001; Rigby &Julian, 2014). Even though acknowledgment is one corner of the "reward triangle" (with authorship and citation)(Costas & Leeuwen, 2012; Cronin & Weaver, 1995), acknowledgment analysis remained relatively underexplored for a long time due to the difficulty of collecting data (Cronin & Shaw, 2007; Cronin et al., 2003; Giles & Councill, 2004; Hyland, 2003). The problem became less of an obstacle when Thomson Reuters' Web of Science (hereinafter WoS) began systematically indexing funding acknowledgment (FA) data from August 2008 forward (Thomson Reuters, 2009). Since then, several researchers use the FA data to track research output, manage funding portfolios, and evaluate the impact of grants (Lewison & Markusova, 2010; Lewison & Roe, 2012; Liu et al., 2015; Rigby, 2011, 2013; Wang & Shapira, 2011; Wang et al., 2012).

Accompanying the burgeoning publications based on acknowledgment analysis, some interesting findings emerge. For example, previous research found 43 % of all WoS indexed publications in 2009 report funding information (Costas & Leeuwen 2012). This percentage is even larger for selected research domains such as molecular biology and biochemistry (Costas & Leeuwen, 2012). The great variances of FA presences among disciplines are also reported by other scholars. Cronin et al. (1993) reported significant differences of acknowledging financial support in four humanities and social sciences disciplines. All publications in *Cell* in selected years include acknowledgments (Cronin & Franks, 2006). At the country level, Wang et al. (2012) reported that among the 10 most prolific countries in Science Citation Index Expanded papers in the year of 2009, over 50 % of Spanish research papers and 70 % of Chinese research papers carried at least one acknowledgment of grant funding. Costas and Leeuwen (2012) also observe that China possesses the largest share of publications acknowledging research funding. Tang and Liu (2015) report that over 90 % of China's highly cited research indexed in the Essential Science Indicators dataset has reported funding agency information.

Meanwhile, concerns about analyzing FA have also been raised. For example, Rigby (2011) explicitly states that there is bias in collecting the FA information, as currently the information is confined to scientific journals only. Lundberg et al. (2006) and Tang (2013) warn that an uncritical use of FA may mislead funding stakeholders and science policy makers. Costas & Leeuwen (2012) cast doubt on the algorithm that Thomson Reuters adopted to index acknowledgment information of research papers. Unfortunately, their concerns and warnings did not incur much attention. Many existing studies utilizing WoS FA information simply neglect these potential problems (Tan et al., 2012; Xu et al., 2015; Zhou & Tian, 2014). No studies up to date have empirically examined the above problems within our best knowledge. This paper aims to advance our understanding of bibliometric analysis using FA by examining potential biases in the WoS practices for collecting and processing FA information. The empirical evidence is provided through both WoS query searching and manual examination of acknowledgment statements. We found that WoS indexing FA information is almost totally dependent on whether or not the paper is indexed in the Science Citation Index Expanded (SCI-E) dataset. FA presence rates vary substantially among non-English papers. In addition, FA information does not report all acknowledgment contents contained in scientific articles.

**Search queries**

WoS includes three searchable field tags that provide funding acknowledgment information: acknowledgment funding organization (FO) identifies funding bodies supporting the research, funding grant (FG) provides grant numbers, and funding text (FT) contains the full text of the authors' acknowledgment section in the paper (Rigby, 2011).

To retrieve a complete set of funding records in WoS, after rounds of trials and errors, we used the following Query #1 searching in the *funding text* field. In order to study data for several full years, we limited our search to publication years 2009 through 2014.[1]

---

[1]All queries were searched by using Web of Science on December 4, 2015.

Query #1 FT=(A* OR B* OR C* OR D* OR E* OR F* OR G* OR H* OR I* OR J* OR K* OR L* OR M* OR N* OR O* OR P* OR Q* OR R* OR S* OR T* OR U* OR V* OR W* OR X* OR Y* OR Z* OR 0* OR 1* OR 2* OR 3* OR 4* OR 5* OR 6* OR 7* OR 8* OR 9*) AND PY=(2009-2014)

Query #1 improves on the search statements used by Wang et al. (2012) and Xu et al. (2015) in the *funding organization* field. Although their query can capture records with any words beginning with any of the 26 letters of the alphabet or the numerals of 0–9 in the *funding organization* field, records with only grant number but no funding organization will not be retrieved. For robustness check, Queries #2 and #3 were also conducted.

Query #2 FO=(A* OR B* OR C* OR D* OR E* OR F* OR G* OR H* OR I* OR J* OR K* OR L* OR M* OR N* OR O* OR P* OR Q* OR R* OR S* OR T* OR U* OR V* OR W* OR X* OR Y* OR Z* OR 0* OR 1* OR 2* OR 3* OR 4* OR 5* OR 6* OR 7* OR 8* OR 9*) AND PY=(2009-2014)

Query #3 FG=(A* OR B* OR C* OR D* OR E* OR F* OR G* OR H* OR I* OR J* OR K* OR L* OR M* OR N* OR O* OR P* OR Q* OR R* OR S* OR T* OR U* OR V* OR W* OR X* OR Y* OR Z* OR 0* OR 1* OR 2* OR 3* OR 4* OR 5* OR 6* OR 7* OR 8* OR 9*) AND PY=(2009-2014)

Our results show that Query #1 (searching the FT field) returned 4,610,481 records, while Query # 2 in FO and Query #3 in FG captured 4,591,259and 3,171,084 records, respectively[2]. We further found that *98 record hits retrieved in FT could not be covered by the combination of* FO or FG (#1 not (#2 or #3)). In comparison, only 4 out of 4,610,387 hits returned by FO or FG are not covered by FT ((#2 or #3) not #1). Thus, unless otherwise specified in this paper, Query #1 searched in the FT field has been used to retrieve WoS funding acknowledgment information.

**Citation index database bias of FA information**

The Web of Science™ Core Collection contains three journal citation databases spanning over 250 disciplines: Science Citation Index Expanded (SCI-E), Social Sciences Citation Index (SSCI), and Arts & Humanities Citation Index (A&HCI).[3] In 2015, 620 SSCI journals and 68 A&HCI journals were also covered by SCI-E.[4]

We applied the Query #1 search to SCI-E, SSCI, and A&HCI separately. Table 1 documents the returned hits and calculated FA presence rates. The coverage biases in the WoS FA information are clearly evidenced by the frequency of funded publications indexed in different citation indices. As shown within the 2009–2014 time band, 9,747,715 publications are indexed in SCI-E, among which 4,608,632 records contain funding acknowledgment with an FA reporting rate of 47 %. This differs from Rigby's (2011) finding that FA information was available only for the papers indexed in SCI (p. 366). We do observe that articles indexed in SSCI and A&HCI also report FA data. But their FA presence rates are extremely low, almost one-third and one-fortieth that of SCI-E.

---
[2] Only three journal citation databases (Science Citation Index Expanded, Social Sciences Citation Index, and Arts & Humanities Citation Index) are included.
[3] For more details please refer to http://wokinfo.com/products_tools/multidisciplinary/webofscience/.
[4] Source : http://ip-science.thomsonreuters.com/mjl/

Table 1 Citation index bias

| Searching set | SCI-E | SSCI | A&HCI | (SSCI or A&HCI) NOT SCI-E |
|---|---|---|---|---|
| Total records (TR) | 9,747,715 | 1,540,644 | 730,918 | 1,504,352 |
| Records with FA | 4,608,632 | 248,856 | 9045 | 2382 |
| Records with FA/TR (%) | 47.28 | 16.15 | 1.24 | 0.16 |

Data source: Thomson Reuters WoS. Time span: 2009-2014.

A further Boolean examination of (SSCI or A&HCI) NOT SCI-E shows that only 2382 out of 1,504,352 records contain FA[5]. The FA reporting rate of WoS papers not indexed in SCI-E dips to 0.16 %. The extremely low FA reporting rate suggests that only FA of publications indexed in SCI-E are systematically recorded. This at least partially accounts for the low FA rates for humanities and social sciences in addition to disciplinary nature and cultural factors suggested by Costas and Leeuwen (2012).

**Languages bias of FA information**

Diaz-Faes and Bordons (2014) reported that WoS captures and processes only FAs that are written in English and that inclusion relies on the assistance provided by Thompson Reuters' technical support team. We are curious whether publications in languages other than English with FAs are also indexed in WoS. We therefore tested the FA presence rate by language of publication.

We applied Query #1 and confined our search to SCI-E. About 9.45 million English publications were identified, and 4.59 million included FAs when all document types considered (Table 2). The FA presence rate of the English-language publications indexed by SCI-E is 49 %. Unlike Diaz-Faes and Bordon's (2014) study, we found that publications written in other languages also have their FA data collected. However, the FA presense rates are extremely low for most other languages. One exception is Chinese—over one-third of Chinese articles also report FA information. Table 2 lists the top 10 languages for the period of 2009–2014 based on the quantity of SCI-E papers. As shown, papers in Chinese, which is the fifth most frequent language in SCI-E, report significantly larger FA presence rate than publications in the more common German, French, and Spanish languages.

---

[5] The following three steps were taken in order to get the records which are indexed by SSCI or A&HCI but not by SCI-E. a) Retrieve records indexed by SSCI or A&HCI (#1); b) Retrieve records indexed by SCI-E (#2); c) Combine sets using the Boolean operator (#1 NOT #2) from the Advanced Search page.

Table 2 Language bias

| Language | Total records | Records with FA | FA presence rate (%) |
|---|---|---|---|
| English | 9,446,993 | 4,592,697 | 48.62 |
| German | 78,616 | 34 | 0.04 |
| French | 48,898 | 32 | 0.07 |
| Spanish | 42,954 | 64 | 0.15 |
| Chinese | 41,743 | 15,246 | 36.52 |
| Portuguese | 32,220 | 44 | 0.14 |
| Polish | 12,225 | 12 | 0.10 |
| Japanese | 8909 | 3 | 0.03 |
| Russian | 7756 | 0 | 0.00 |
| Turkish | 6265 | 5 | 0.08 |

Data source: Thomson Reuters WoS-SCIE. Time span: 2009-2014.

**Acknowledgment type bias of FA information**

Researchers acknowledge support in their paper for a variety of reasons. Previous studies have categorized acknowledgments into different types: moral support; financial support; access to facilities, data, etc.; clerical support; technical support; and peer interactive communication (Cronin, 1991; Cronin, McKenzie, & Rubio, 1993). The WoS name for this field *funding text* intuitively delivers the message that the acknowledgment is about financial support of the research. Yet it remains unclear if all types of acknowledgment are systematically collected in WoS. No research has examined this issue with one exception: Costas & Leeuwen (2012) manually checked the acknowledgments of their own publications and found that WoS did not include the acknowledgment texts of papers that did not contain funding acknowledgment (p. 1650). To explore this question, we chose a journal and downloaded the full texts of all its articles published in 2014 and manually examined the acknowledgment sections. Following the common practice of selecting top-ranking journals suggested by previous studies (Connor, 2004; Bazerman, 1994; Cronin, McKenzie, & Rubio, 1993), we purposely chose *Journal of the Association for Information Science and Technology* (JASIST)[6], a leading journal in library information science (LIS) and computer science indexed in both SCI-E and SSCI. Our manual analysis shows that 215 papers[7] were published in JASIST in 2014, with 116 containing acknowledgments sections. We applied Query #1 in WoS but restricted searching to JASIST in the year of 2014, and the search returned only 83 hits. The left are 33 JASIST acknowledgment-bearing articles which could not be retrieved by searching the FT field in WoS. Without exception, these articles' acknowledgment sections do not contain research funding information. This finding provides further evidence supporting the claim of Costas and Leeuwen (2012) that only acknowledgments with funding information are collected in WoS.

---

[6]JASIST was changed to its current name from *Journal of the American Society for Information Science and Technology* in 2014.
[7]These 215 papers consist of 183 original articles, 15 book reviews, 8 letters, 2 reviews, 2 editorial materials, 4 biographical-Item, and one correction.

**Conclusion and Discussion**

Funding acknowledgment (FA) is an increasingly institutionalized practice across scientific fields. Previous studies have proposed cautions regarding FA analysis: misspellings and variants of funding organizations' names (Wang & Shapira, 2011; Lewison & Roe, 2012; Tang, 2013), ghost and gift funding organizations (Claxton, 2005; Giles & Councill, 2004), and unconsciously over- or under-reported financial supporting information (Tang et al., 2015; Costas & Leeuwen, 2012). In addition to those pitfalls, the inherent biases in Thomson Reuters' practices for collecting FA information should also be clear for future research.

This study provides empirical evidence of the limitations in WoS FA information collection. We found that the WoS database records an acknowledgment only if it contains funding information, and thus it is not recommended for analyzing other types of acknowledgment without complementary information. For WoS databases, only FAs in journals indexed by SCI-E are systematically recorded. In other words, the WoS FA data is not suitable for analyzing social science and humanities research. In addition, WoS records FA information almost exclusively for papers in English and for those in Chinese with English FAs, so the data is not recommended for analyzing publications written in languages other than English and Chinese.

To conclude, although FA analysis opens a wide range of possibilities for linking scientific input and output (such as the correlation between funding with collaboration and research performance), we argue that the pitfalls and potential impacts on the results of bibliometric analysis of FA should be taken into account when undertaking this type of analysis. These caveats are particularly important when using bibliometric analysis to make comparisons across different countries and research disciplines.

**Acknowledgments:** This research is supported by National Natural Science Foundation of China (#71303147 and #71132006) and Shanghai Soft Science Program (#14692102900). We appreciate the valuable comments and feedback of two anonymous reviewers. The views expressed herein are those of the authors and not necessarily those of the funders listed here. We are responsible for any errors.